\documentclass{article}
\usepackage[utf8]{inputenc}
\usepackage{authblk}
\usepackage[dvips]{graphicx}
\graphicspath{{noiseimages/}}
\usepackage{xcolor}
\usepackage[T2A]{fontenc}
\usepackage{tikz}
\usepackage{pgfplots}
\pgfplotsset{compat=1.7}
\usetikzlibrary{intersections, pgfplots.fillbetween}
\usetikzlibrary{decorations.pathmorphing}
\usepackage[mode=buildnew]{standalone}
\usepackage[toc,page]{appendix}
\usepackage{amsmath}
\usepackage{amssymb}
\usepackage{float}
\usepackage{comment}
\usepackage{a4wide}
\usepackage[english]{babel}
\usepackage{hyperref}
\definecolor{urlcolor}{HTML}{990000}
\definecolor{linkcolor}{HTML}{005F5F} 
\hypersetup{pdfstartview=FitH,  linkcolor=linkcolor,urlcolor=urlcolor, colorlinks=true,citecolor=blue}
\usepackage[page]{appendix}

\newcommand{\w}{\omega}

\author[1,2]{P.A.Anempodistov\footnote{\tt anempodistov.pa@phystech.edu }}

\affil[1]{Moscow Institute of Physics and Technology, Institutskii per. 9, 141700, Dolgoprudny, Russia}
\affil[2]{ Institute for Theoretical and Experimental Physics, B. Cheremushkinskaya 25, 117218, Moscow, Russia}

\title{\textcolor{black}{Remarks on the thermofield double state in 4D black hole background}}
\begin{document}

\numberwithin{equation}{section}

\maketitle

\begin{abstract}
Recently it was shown that there is an anomalous singularity of propagators in spacetimes with horizons for thermal states with non--canonical temperatures. In this paper we extend these observations to the situation when the background geometry is given by the four--dimensional Schwarzschild and Reissner-Nordström black holes. 
Namely, we demonstrate that the two-point function in the free scalar field theory acquires anomalous singularity when the two points are located on the horizon. This singularity is anomalous in the sense that the coefficient before the divergent term in the two-point function differs from the canonical one and depends explicitly on the temperature of the state. As it was previously shown, it leads to the explosive behavior of the regularised stress-energy tensor on the horizon when the temperature of the state does not coincide with the canonical temperature of the horizon.

\end{abstract}

\newpage


\newpage

\section{Introduction }
When discussing quantum field theory in the eternal Schwarzschild black hole background, one usually considers the Boulware, Unruh, or Hartle-Hawking states \cite{Boulware:1974dm}, \cite{Unruh:1976db}, \cite{Hartle:1976tp}, \cite{DeWitt:2003pm}, \cite{Candelas:1980zt}. While the Boulware state is defined by taking the exact modes to be positive--definite w.r.t. $\partial / \partial t$ Killing vector, which would give the usual Poincare--invariant ground state in flat--spacetime QFT, the Unruh and Hartle-Hawking states are defined by introducing Kruskal coordinates (canonical affine parameters on the past and future horizons) and taking modes to be positive--definite w.r.t. these Kruskal coordinates. The outcome of the two latter definitions of positive-definite modes is that either both or one of the out--going or in--going modes are in a thermal state with the temperature that is equal to the Hawking one. Which state is more suitable to a certain physical situation is analyzed by the calculation of the energy-momentum tensor on the horizon and at spatial infinity \cite{Candelas:1980zt}, \cite{Akhmedov:2015xwa}. However, there was no consideration of the situation in which both out--going and in--going modes are in thermal states with temperatures which generically do not coincide and which are not equal to the Hawking temperature. It would be natural to assume that this state is the best approximation for a physical situation when one considers a black hole in a box of a gas with a temperature which is not equal to the Hawking one. The same question can be posed analogously in the Rindler and de Sitter space--times with respective canonical temperatures of the horizons. 

The de Sitter space, Rindler space, and two-dimensional black hole examples were analysed in \cite{Akhmedov:2020ryq}. It was shown that when one considers fields in a thermal state with temperature which is not equal to the canonical temperature of the horizon of the background geometry, then propagators, both points of which are located at the horizon, acquire an anomalous singularity. This singularity is anomalous in a sense that the standard light--like separation divergence comes from high frequencies and does not depend on the temperature of the state, while at the horizon the singularity comes from the infrared region in the integral over frequencies and there is an explicit dependence on the temperature of the state in which the propagator is evaluated. 

We consider the following quantum field theory
\begin{align*}
    S=\frac{1}{2}\int d^4 x \sqrt{-g}\Big[ \partial_\mu \phi \partial^\mu \phi -\mu^2 \phi^2\Big]
\end{align*}
on the Schwarzschild and Reissner-Nordström black hole backgrounds.


In \cite{Akhmedov:2020ryq} only the two--dimensional analogue of the Schwarzschild black hole was considered. In this paper these results are extended to the four--dimensional black holes. We start in the section \ref{modes_wight_sch} by constructing modes and the Wightman function in the Schwarzschild black hole background. In the section \ref{canonical_sing} it is shown that the singularity of the Wightman function is canonical when its points are light-like separated outside the horizon, i.e. the divergent term in the Wightman function has the standard coefficient. Then, in the section \ref{anom_sing} we demonstrate that the Wightman function acquires anomalous singularity when both of its points are located on the horizon, i.e. the coefficient before the divergent term differs from the canonical one and explicitly depends on the temperature of the out-going modes. Finally, we conclude in section \ref{RNsection} by showing that the anomalous singularity of the Wightman function also occurs in the Reissner-Nordström black hole background.

\section{Schwarzschild black hole}

In this section we consider both massive and massless fields in Schwarzschild spacetime. First, we set the notations and define modes and thermal Wightman functions. For more detailed review the reader may refer to \cite{Boulware:1974dm}, \cite{Unruh:1976db}, \cite{Hartle:1976tp}, \cite{DeWitt:2003pm}, \cite{Candelas:1980zt}. Then, the Wightman function is computed when both of its points are located at the event horizon.

\subsection{Modes and Wightman function} \label{modes_wight_sch}
\begin{figure}[!h]
\centering
\includestandalone[width=0.7\textwidth]{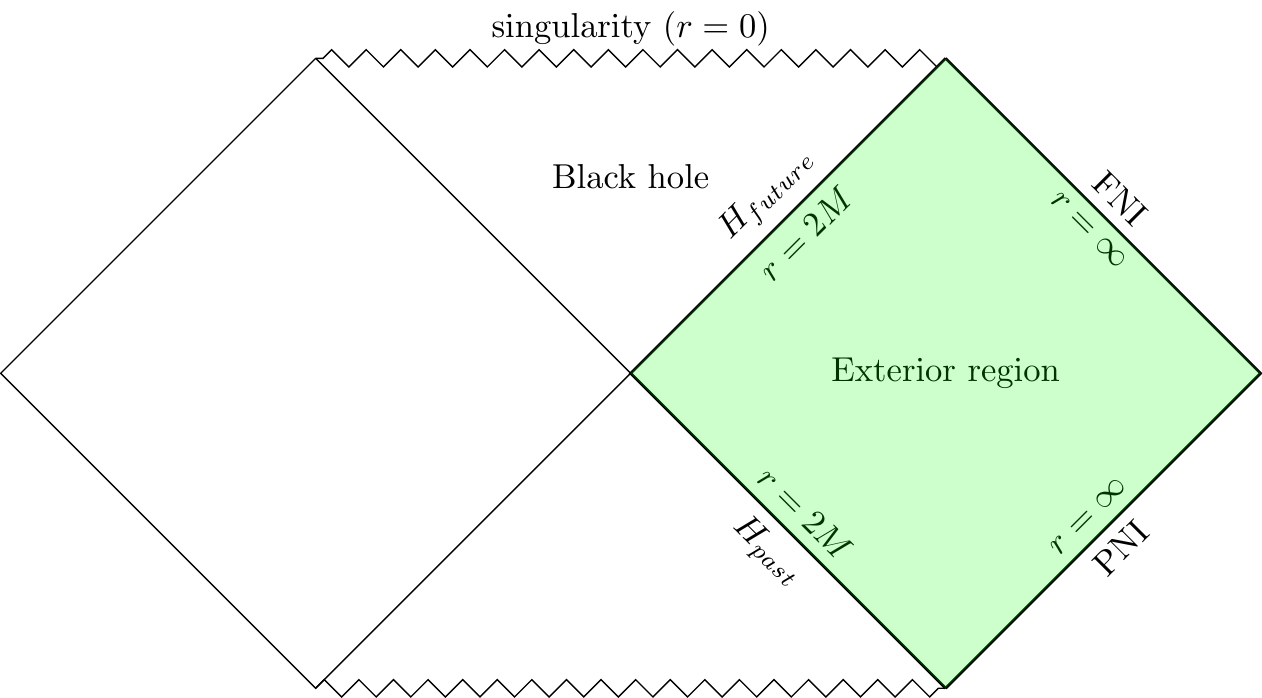}
\caption{Penrose diagram of the Schwarzschild black hole. Here and further "PNI" and "FNI" stand for "past null infinity" and "future null infinity" correspondingly.} \label{picschw}
\end{figure}

The Schwarzschild black hole background is given by the metric:
\begin{align}
    &ds^2 = \bigg( 1-\frac{2M}{r} \bigg) dt^2 - \frac{dr^2}{1-\frac{2M}{r}}-r^2(d \theta^2 + \sin^2 \theta d \varphi^2).
\end{align}
Here Eddington-Finkelstein coordinates are defined as:
\begin{align}
    &u = t- r_*, \qquad v=t+r_*, \nonumber \\
    &r_* = r+2M \log \big( r/2M-1 \big).
\end{align}
In Sec. \ref{canonical_sing} it will be shown that the Wightman function has a standard divergence when its points are lightlike separated near the horizon. Then, in Sec. \ref{anom_sing} I will consider the situation in which both points of the Wightman function are located on the future horizon, which is denoted as $H_{future}$ in Fig. \ref{picschw}.

\subsubsection{Massless case}
Solutions of the corresponding d'Alembertian equation are as follows:
\begin{equation} \label{split}
    u_{\w lm} (x) = (4 \pi |\omega|)^{-1/2} r^{-1} f_{\w l} (r,t) Y_{lm} (\Omega),
\end{equation}
where $\Omega$ denotes the angular coordinates. 
To define the set of Boulware modes, we impose the condition that the modes are positive-definite w.r.t. ${\partial}/{\partial t}$ Killing vector, so one can represent the function $f_{\w l} (r,t)$ in the form
\begin{equation} \label{f_split}
    {f}_{\w l} (t,r) = e^{-i|\w| t} {R}_l (\w | r) ,
\end{equation}
in which the radial function $R_l (\w | r)$ solves the equation
\begin{align} \label{radial}
    &\frac{d^2R}{dr_*^2}+\big[ \w^2 - V_l (r) \big] R = 0,
\end{align}
with the potential
\begin{equation} \label{potential}
    V_l (r) = \bigg( 1 - \frac{2M}{r} \bigg) \bigg(  \frac{l(l+1)}{r^2} + \frac{2M}{r^3} \bigg).
\end{equation}
The complete set of modes in the outer region of the Schwarzschild black hole is given by the out-going and in-going modes, denoted accordingly as:
\begin{eqnarray} \label{set1}
&\overrightarrow{u}_{\omega lm} (x) = (4 \pi |\omega|)^{-1/2} e^{-i |\omega| t} r^{-1} \overrightarrow{R}_l(\omega | r) Y_{lm} (\theta, \varphi), \nonumber \\
&\overleftarrow{u}_{\omega lm} (x) = (4 \pi |\omega|)^{-1/2} e^{-i |\omega| t} r^{-1} \overleftarrow{R}_l(\omega | r) Y_{lm} (\theta, \varphi),
\end{eqnarray}
where the radial functions satisfy the boundary conditions as follows
\begin{equation}
                \overrightarrow{R}_l(\omega| r) = \begin{cases}
                 e^{i \omega r_*} + \overrightarrow{A}_l(\omega) e^{-i \omega r_*},
                & r \rightarrow 2M \\
                {B}_l (\omega) e^{i \w r_*}, & r \rightarrow \infty 
                \end{cases} \nonumber
            \end{equation}
    and
    \begin{equation} \label{boundary_cond}
                \overleftarrow{R}_l(\omega |r) = \begin{cases}
                {B}_l ({\omega}) e^{ - i \omega r_*}, & r \rightarrow 2M \\
                e^{-i \w r_*} + \overleftarrow{A}_l ({\omega}) e^{i \w r_*}, & r \rightarrow \infty
                \end{cases}
            \end{equation}
As we have only two types of modes, the mode decomposition of the field operator is as follows
\begin{equation}
    \phi (x) = \sum_{lm} \int_0^{\infty} d \w \bigg( a_{\w lm} \overrightarrow{u}_{\w lm} (x) + b_{\w lm} \overleftarrow{u}_{\w lm} (x) + h.c. \bigg).
\end{equation}
Then for the Wightman function in a generic state with zero anomalous quantum averages (i.e. with $\langle a_{p} a_{q}\rangle = \langle a^{\dagger}_{p} a^{\dagger}_{q}\rangle =\langle b_{p} b_{q}\rangle = \langle b^{\dagger}_{p} b^{\dagger}_{q}\rangle = 0 $), but which is not necessary Fock ground state, one has:

\begin{multline} \label{wightman_general}
    W (x,x') \equiv \langle \phi (x) \phi(x') \rangle
= \int^{+\infty}_{0} {d \omega} \int^{+\infty}_{0} {d \omega'} \bigg[ \langle a_{\omega} a_{\omega'}^{\dagger} \rangle \overrightarrow{u}_{\omega}(x) \overrightarrow{u}^*_{\omega'}(x')
 + \langle a_{\omega}^{\dagger} a_{\omega'} \rangle \overrightarrow{u}^*_{\omega}(x) \overrightarrow{u}_{\omega'}(x')
 + \\
 +\langle b_{\omega} b_{\omega'}^{\dagger} \rangle\overleftarrow{u}_{\omega}(x) \overleftarrow{u}^*_{\omega'}(x') + \langle b_{\omega}^{\dagger} b_{\omega'} \rangle
  \overleftarrow{u}^*_{\omega}(x) \overleftarrow{u}_{\omega'}(x') \bigg].
\end{multline}
Here we have omitted spherical harmonic indices to simplify the equation.
Allowing the temperatures of out--going and in--going modes to differ from each other, in this paper we will study states in which
\begin{equation}
     \langle a_{\omega}^{\dagger} a_{\omega'} \rangle = \frac{ \delta(\w -\w')}{e^{\beta_R |\w|}-1}, \qquad \langle b_{\omega}^{\dagger} b_{\omega'} \rangle = \frac{\delta(\w -\w')}{e^{\beta_L |\w|}-1}, 
\end{equation}
so the Wightman function can be written as
\begin{multline} 
    W(x,x') = \sum_{lm} \int^{+\infty}_{0} {d \omega} \bigg[  \frac{\overrightarrow{u}_{\omega lm}(x) \overrightarrow{u}^*_{\omega lm}(x')}{1-e^{-\beta_R |\w|}} 
 + \frac{\overrightarrow{u}^*_{\omega lm}(x) \overrightarrow{u}_{\omega lm}(x')}{e^{\beta_R |\w|}-1} + 
  \frac{\overleftarrow{u}_{\omega lm}(x) \overleftarrow{u}^*_{\omega lm}(x')}{1-e^{-\beta_L |\w|}}+ \\ + \frac{\overleftarrow{u}^*_{\omega lm}(x) \overleftarrow{u}_{\omega lm}(x')}{e^{\beta_L |\w|}-1} \bigg] = \\
  =\sum_{lm} \int^{+\infty}_{0} \frac{d \omega}{4\pi |\w|} \frac{1}{rr'}\bigg[ \frac{e^{i |\w| (t-t')} \overrightarrow{R}^*_l (\w|r) \overrightarrow{R}_l (\w|r') }{e^{\beta_R | \w|}-1} 
  +\frac{e^{i |\w| (t-t')} \overleftarrow{R}^*_l (\w|r) \overleftarrow{R}_l (\w|r') }{e^{\beta_L |\w|}-1} \bigg] Y_{lm} (\Omega) Y^*_{lm} (\Omega') + \\+
  \sum_{lm} \int^{0}_{-\infty} \frac{d \omega}{4\pi |\w|} \frac{1}{rr'}\bigg[ \frac{e^{-i |\w| (t-t')} \overrightarrow{R}^*_l (\w|r) \overrightarrow{R}_l (\w|r') }{1-e^{-\beta_R | \w|}}
  +\frac{e^{-i |\w| (t-t')} \overleftarrow{R}^*_l (\w|r) \overleftarrow{R}_l (\w|r') }{1-e^{-\beta_L |\w|}} \bigg] Y_{lm} (\Omega) Y^*_{lm} (\Omega'),
\end{multline}
where in the last relation we have made the substitution $\w \to -\w$ for the first and the third terms in the first line. Also we have used the relation $\overrightarrow{R}_l(-\w|r) = \overrightarrow{R}^*_l(\w|r)$, which can be deduced from \eqref{near_horizon_mode} or \eqref{pt_reflection} with \eqref{boundary_cond}. Then we put $|\w| = +\w$ in the first integral of the last equality and $|\w| = -\w$ in the second integral to obtain the following form of the Wightman function:
\begin{multline} \label{wightman}
    W(x,x') = \sum_{lm} \int^{+\infty}_{-\infty} \frac{d \omega}{4\pi \w} \frac{1}{rr'}\bigg[ \frac{e^{i \w (t-t')} \overrightarrow{R}^*_l (\w|r) \overrightarrow{R}_l (\w|r') }{e^{\beta_R  \w}-1}
  +\\+\frac{e^{i \w (t-t')} \overleftarrow{R}^*_l (\w|r) \overleftarrow{R}_l (\w|r') }{e^{\beta_L \w}-1} \bigg] Y_{lm} (\Omega) Y^*_{lm} (\Omega').
\end{multline}
For the case $\beta_R = \beta_L = \infty$ we obtain the Boulware state, while $\beta_R=\frac{2\pi}{\kappa}, \beta_L = \infty$ corresponds to the Unruh state, and $\beta_R=\beta_L=\frac{2\pi}{\kappa}$ results in the Hartle-Hawking state. Note that in the limit $\beta=\infty$ Bose-Einstein distribution reduces to Heaviside theta-function $\theta(-\w)$ and the Wightman function reduces to what one would have obtained from \eqref{wightman_general} by taking the state to be the Fock vacuum. Here $\kappa = (4M)^{-1}$ is the surface gravity of the black hole. 
\subsubsection{Massive case}

In the case of massive field ($\mu \neq 0$), the equation for the radial function is
\begin{align} \label{radial}
    &\frac{d^2R}{dr_*^2}+\big[ \w^2 - V_l (r) \big] R = 0,
\end{align}
with the potential
\begin{equation} \label{potential}
    V_l (r) = \bigg( 1 - \frac{2M}{r} \bigg) \bigg(  \frac{l(l+1)}{r^2} + \frac{2M}{r^3} + \mu^2 \bigg),
\end{equation}
which tends to $\mu^2$ in the $r^* \to \infty$ limit. So as in \cite{Akhmedov:2020ryq}, \cite{Akhmedov:2015xwa}, \cite{Akhmedov:2016uha} there are modes with $\omega^2 < \mu^2$ that are localised near the horizon and which are exponentially decreasing in the classically inaccessible domain. 
Rewriting the equation \eqref{radial} in terms of coordinate $r$, taking the near-horizon limit and introducing the new coordinate $\xi^2 = \frac{r}{2M}-1$ leads to the equation
\begin{equation}
    \frac{d^2 R}{d\xi^2} + \frac{1}{\xi} \frac{dR}{d\xi} + \bigg[ \frac{(4M\w)^2}{\xi^2} -4l(l+1)-(4M\mu)^2 \bigg] R = 0,
\end{equation}
solutions of which are given by the modified Bessel functions of the second kind. Keeping in mind the condition of exponential decrease in the classically inaccessible domain, one obtains
\begin{equation} \label{radial_massive_sch}
    R \approx  \sqrt{\frac{\w \sinh({4\pi M \w})}{\pi M}} K_{4iM\w} \big( 2 \sqrt{(2M\mu)^2+l^2+l} \, \xi \big).
\end{equation}
Then, the modes are denoted as
\begin{align}
    &\varphi_{\w lm} (x) = \frac{1}{\sqrt{ \pi |\w|}} e^{-i |\w| t} R_{ l} (\w | r) Y_{lm} (\Omega),
\end{align}
where the radial part is given by \eqref{radial_massive_sch}, so that near the horizon one has
\begin{gather} \label{delta_sch}
    R_{l} (\w | r) \approx \frac{1}{r} \cos (\w r_* + \delta_{\w l}), \\
    \delta_{\w l} \approx \frac{\pi}{2} +2M\w \log (4M^2 \mu^2 +l^2+l) - 2M\w - \text{arg} \, \Gamma(1+4iM\w).
\end{gather}
The modes with $\w^2>\mu^2$ are similar to the massless case:
\begin{align} \label{f_modes}
    &\overrightarrow{F}_{\omega lm} (x) = (4 \pi |\omega|)^{-1/2} e^{-i |\omega| t} r^{-1} \overrightarrow{F}_l(\omega | r) Y_{lm} (\Omega), \nonumber \\
&\overleftarrow{F}_{\omega lm} (x) = (4 \pi |\omega|)^{-1/2} e^{-i |\omega| t} r^{-1} \overleftarrow{F}_l(\omega | r) Y_{lm} (\Omega),
\end{align}
where their radial parts obey the following boundary conditions:
\begin{align}
                &\overrightarrow{F}_l ({\omega} | r) = \begin{cases}
                e^{i \omega r_*} + \overrightarrow{C}_{\omega} e^{-i \omega r_*}, 
                & \qquad r \rightarrow 2M \\
                \sqrt{\frac{\omega}{p}} {D}_{\omega} e^{i p r_*}, & \qquad r \rightarrow \infty 
                \end{cases} \\
                &\overleftarrow{F}_l ({\omega} |r) = \begin{cases}
                {D}_{\omega} e^{ - i \omega r_*}, & r \rightarrow 2M \\
                \sqrt{\frac{\omega}{p}} [ e^{-ipr_*} + \overleftarrow{C}_{\omega} e^{ipr_*}], & r \rightarrow \infty 
                \end{cases}
            \end{align}
            with $p=\text{sgn} (\w) \cdot \sqrt{\omega^2 - \mu^2}$, \, $\w^2 > \mu^2$.

The mode decomposition of the field operator for the massive case is as follows
\begin{multline} \label{mode_massive_sch}
    \phi(x) = \sum_{lm} \int_0^{\mu} d\w \big( \alpha_{\w lm} \varphi_{\w lm} + \alpha_{\w lm}^{\dagger} \varphi_{\w lm}^* \big) + \\
    +\sum_{lm} \int_{\mu}^{\infty} d \w \big( \beta_{\w lm} \overrightarrow{F}_{\w lm} + \beta_{\w lm}^{\dagger} \overrightarrow{F}^*_{\w lm} + \gamma_{\w lm} \overleftarrow{F}_{\w lm} + \gamma_{\w lm}^{\dagger} \overleftarrow{F}_{\w lm}^* \big),
\end{multline}
and the Wightman function has the following form:
\begin{multline}
    W(x,x') = \sum_{lm} \int^{+\mu}_{-\mu} \frac{d \omega}{\pi \w}  \frac{e^{i \w (t-t')} R^*_l(\w|r) R_l (\w|r') }{e^{\beta_0 \w}-1} Y_{lm}(\Omega) Y^*_{lm}(\Omega') + \\
    +\sum_{lm} \int_{|\w|>\mu} \frac{d \omega}{4\pi \w} \frac{1}{rr'} \bigg[ \frac{ e^{i \w (t-t')} \overrightarrow{F}^*_{ l}(\w | r) \overrightarrow{F}_{l}(\w|r')}{e^{\beta_R \w}-1} 
 +
  \frac{ e^{i \w (t-t')} \overleftarrow{F}^*_{l}(\w|r) \overleftarrow{F}_{l}(\w|r')}{e^{\beta_L \w}-1} \bigg] Y_{lm}(\Omega) Y^*_{lm}(\Omega') .
\end{multline}
Note that we have added here the inverse temperature $\beta_0$ for the modes with $\w^2 \leq \mu^2$, which does not have to be equal neither to $\beta_R$ nor to $\beta_L$.

\subsection{Canonical singularity} \label{canonical_sing}
First we want to explicitly show in what sense the divergences derived further in this section are anomalous. To do so, we consider the Wightman function \eqref{wightman} when both of it's points are located in the vicinity of the horizon, but not exactly on it. We will take light-like separation of these two points, keeping their angular coordinates equal. We restrict ourselves to the massless fields in this subsection.

The metric near the horizon can be written as
\begin{equation}
    ds^2 \approx \rho^2 d \eta^2 - d \rho^2 -r^2(\rho) \, d\Omega^2,
\end{equation}
where
\begin{align}
    \rho = 4M \sqrt{\frac{r}{2M}-1} ,\qquad \eta = \frac{t}{4M}.
\end{align}
Then, the square of the geodesic distance between the two points under consideration is written as
\begin{equation} \label{geodesic}
    L^2 \approx 2 \rho_1 \rho_2 \cosh (\eta_1-\eta_2) - \rho_1^2-\rho_2^2.
\end{equation}
 One can expand the potential near the horizon and find a solution (in terms of modified Bessel function of the second kind) that satisfies boundary condition \eqref{boundary_cond}:

\begin{equation} \label{near_horizon_mode}
    \overrightarrow{R}_l (\w|r) \approx \frac{2}{\Gamma (-i 4 M \w)} e^{-2iM \w \log (l^2+l+1)+2iM\w} K_{4iM\w} (2 \sqrt{l^2+l+1} e^{\frac{r_*-2M}{4M}}),
\end{equation}
from which one can also find that
\begin{equation}
    \overrightarrow{A}_l (\w) \approx - e^{4iM\w - 4iM\w \log(l^2+l+1) + 2i \text{arg} \Gamma(1+4iM \w) }.
\end{equation}

Now, plugging \eqref{near_horizon_mode} into \eqref{wightman}, one obtains that
\begin{multline} \label{prop_non_horizon}
    W(x,x') \approx \frac{1}{\pi^2} \frac{1}{(2M)^2} \sum_{lm} Y_{lm} (\Omega) Y_{lm}^* (\Omega) \int^{\infty}_{-\infty} d\w \frac{e^{\frac{i \w (t-t')}{4M}}}{e^{\frac{\beta_R \w}{4M}}-1} \sinh( \pi \w) \cdot \\ \cdot
    K_{i\w} (2 \sqrt{l^2+l+1} e^{\frac{r_*-2M}{4M}}) K_{i\w} (2 \sqrt{l^2+l+1} e^{\frac{r_*'-2M}{4M}}).
\end{multline}
The simple formula is available if we take $\beta_R = \frac{8\pi M}{n}$ with $n$ being integer \cite{Akhmedov:2019esv}, \cite{Akhmedov:2020qxd}. Namely, we employ the fact that in this case
\begin{equation}
    \frac{\sinh(\pi \w)}{e^{\frac{\beta_R \w}{4M}}-1} = \frac{e^{-\pi \w}}{2} \sum_{k=1}^{n} e^{\frac{2\pi \w (k-1)}{n}}.
\end{equation}
Plugging this expressions into \eqref{prop_non_horizon} and evaluating the integral over $\w$, we obtain that
\begin{multline} \label{wightman_discrete_sum}
    W(x,x') \approx \frac{1}{2 \pi} \frac{1}{(2M)^2} \sum_l \frac{(2l+1)}{4\pi} \times \\ \times
    \sum_{k=1}^{n} K_0 \bigg( 2 \Tilde{l} \sqrt{e^{\frac{r_*-2M}{2M}} + e^{\frac{r_*'-2M}{2M}} +2 e^{\frac{r_*-2M}{4M}} e^{\frac{r_*'-2M}{4M}} \cosh \bigg[ -\frac{t-t'}{4M} +i \pi - i \frac{2\pi (k-1)}{n}  \bigg] } \bigg),
\end{multline}
where $\Tilde{l} = \sqrt{l^2+l+1}$.
Here only the $k=1$ term depends on the geodesic distance between the two points \eqref{geodesic}, so detaching this terms from the sum and writing the Wightman function as:
\begin{equation} \label{wightman_k}
    W(x,x') = W_{k = 1} (x,x') + W_{k \neq 1} (x,x'),
\end{equation}
we obtain that for a light-like separated $x$ and $x'$, i.e. with $L=0$ (but not yet sitting at the horizon):
\begin{multline}
    W_{k=1}(x,x') \approx \frac{1}{2\pi} \frac{1}{(2M)^2} \sum_l \frac{(2l+1)}{4\pi}
    K_0 \bigg( \frac{\sqrt{l^2+l+1}}{2M} \sqrt{-{L^2}} \bigg) \approx \\
    \approx \frac{1}{4\pi^2} \frac{1}{(2M)^2} \int_0^{\infty} dl \, l K_0 \bigg( \frac{l}{2M} \sqrt{-L^2} \bigg) = -\frac{1}{4 \pi^2 L^2},
\end{multline}
and as the terms in $W_{k \neq 1} (x,x')$ do not depend on the geodesic distance between $x$ and $x'$, they are negligible when we are considering light-like separation of the points out of the horizon. Hence, we get that
\begin{equation}
    W(x,x') \approx  -\frac{1}{4 \pi^2 L^2},
\end{equation}
i.e. we have the standard coefficient of $(4\pi^2)^{-1}$ in front of the divergence.

However, when the points are located on the horizon, in \eqref{wightman_k} $k \neq 1$ terms also have divergent contributions. We can place the two points on the future horizon by taking the following limit:
\begin{align} \label{limit}
    &t=-\lambda, \qquad r_*=\lambda, \nonumber \\
    &t'=-\lambda, \qquad r_*'=\lambda+c, \qquad \lambda \to -\infty.
\end{align}
Now we demonstrate that in this limit the Wightman function acquires additional divergent terms. In the case when $n=2$, the Wightman function has the following form:
\begin{multline}
    W(x,x') \approx \frac{1}{2\pi} \frac{1}{(2M)^2} \sum_l \frac{(2l+1)}{4\pi} \bigg\{
    K_0 \bigg( \frac{\sqrt{l^2+l+1}}{2M} \sqrt{-{L^2}} \bigg) + \\ + K_0 \bigg( \frac{\sqrt{l^2+l+1}}{2M} \sqrt{-{L^2 \Lambda}} \bigg) \bigg\} \approx -\frac{1}{4 \pi^2 L^2} \frac{1+\Lambda}{\Lambda},
\end{multline}
where we have expressed the argument of the second term in the curly bracket through the geodesic distance in the limit \eqref{limit} via the coefficient $\Lambda$, which has the form
\begin{equation}
    \Lambda = \frac{1+e^{c/2M}+2 e^{c/4M}}{1+e^{c/2M}-2e^{c/4M}}.
\end{equation}
However, in this approach it is not clear how to obtain the answer for the Wightman function even in the case of the discrete inverse temperature $\beta_R = \frac{8\pi M}{n}$ with an arbitrary $n$, let alone the case of the arbitrary $\beta_R$. This difficulty comes from the fact that as opposed to the static patch of de Sitter space, in this case one cannot express the Wightman function with the non-canonical temperature as the sum of the canonical temperature Wightman functions. In the following section we calculate the Wightman function using the asymptotic form of the modes near the horizon, and the result is that the coefficient of the singularity of the Wightman function depends explicitly on the temperature (as it is shown in  \eqref{sing_schw_massive} and \eqref{schw_res} below). Furthermore, by taking high-frequency limit in \eqref{prop_non_horizon}, it can be shown that the singularity outside the horizon is an UV effect, while at the horizon the singularity comes from the IR region of the frequencies. This justifies the fact that everywhere outside the horizon the divergence does not depend on the temperature, while at the horzion there is an explicit dependence on the temperature (i.e. on the low laying state).

\subsection{Anomalous singularity} \label{anom_sing}

In this section the anomalous singularity in the Wightman function on the horizon for generic thermal state is derived. 

\subsubsection{Massless case}

Using the asymptotic behavior of the modes \eqref{boundary_cond} in the limit \eqref{limit},
the Wightman function \eqref{wightman} can be written as:
\begin{multline} \label{prop}
    W(x,x') \approx \frac{1}{(2M)^2} \sum_{lm} Y^*_{lm} (\Omega) Y_{lm} (\Omega')  \int_{-\infty}^{\infty} \frac{d \w }{4 \pi \w} \bigg[ \frac{1}{e^{\beta_R \w }-1} \bigg( e^{-i\w c} + \overrightarrow{A}_l (\w) e^{-2 i \lambda \w } + \\ + \overrightarrow{A}_l^* (\w) e^{2 i \lambda \w } + | \overrightarrow{A}_l (\w) |^2 e^{i\w c} \bigg) + \frac{1}{e^{\beta_L \w}-1} | B_l (\w) |^2 e^{i \w c} \bigg].
\end{multline}
To integrate over $\w$, analytic expression for reflection coefficient $\overrightarrow{A}_l (\w)$ is needed. To compute it, one needs to know transition coefficients between solutions of confluent Heun equation near the horizon and infinity. However, there is no knowledge about these coefficients \cite{Fiziev:2005ki}, but we can approximate potential for radial function by P{\"o}schl-Teller potential \cite{Chirenti:2017mwe},
\cite{Berti:2009kk}, which has the form
\begin{equation}
    \Tilde{V}_l(r) = \frac{V_0}{\cosh^2 \alpha (r_* - \Tilde{r}_*) },
\end{equation}
where $\Tilde{r}_*$ is the position of the peak and
\begin{align}
    V_0 = V_l (\Tilde{r}_*), \qquad \alpha^2 = - \frac{1}{2V_0} \frac{d^2 V_l}{dr_*^2}_{|r_*=\Tilde{r}_*}.
\end{align}
In this potential we can easily compute the reflection coefficient $\overrightarrow{A}_l (\w)$, which is found to be
\begin{equation} \label{pt_reflection}
    \overrightarrow{A}_l (\w) = \frac{\Gamma({i \w}/{\alpha}) \Gamma(-i {\w}/{\alpha}-s) \Gamma(1+s-i {\w}/{\alpha})}{\Gamma(-i {\w}/{\alpha}) \Gamma(1+s) \Gamma(-s) }, \qquad s = -\frac{1}{2} + \frac{i}{2} \sqrt{\frac{4V_0}{\alpha^2}-1}.
\end{equation}
So we have to evaluate the integral 
\begin{multline}
    \int_{-\infty}^{\infty} \frac{d \w }{4 \pi \w} \frac{ \overrightarrow{A}_l (\w) e^{-2 i \lambda \w }}{e^{\beta_R \w}-1} = \int_{-\infty}^{\infty} \frac{dz}{4 \pi z} \frac{1}{e^z-1}  \frac{\Gamma(iz/\alpha \beta_R) \Gamma(-iz/ \alpha \beta_R-s) \Gamma(1+s-iz/ \alpha \beta_R)}{\Gamma(1+s) \Gamma(-s) \Gamma(-iz/ \alpha \beta_R)} e^{2 i \frac{| \lambda |}{\beta_R} z} = \\
    = \pi i \text{Res}(f(z),0) + 2\pi i \text{Res}(f(z), \overline{0}),
\end{multline}
where the second term $\text{Res}(f(z), \overline{0})$ denotes the sum over all residues on the upper half-plane poles.

The residue at $z=0$ in the limit $\lambda \to -\infty$ is computed to be
\begin{equation}
    \pi i \text{Res}(f(z),0) =  \frac{|\lambda|}{2 \beta_R}.
\end{equation}
Here an additional term of the form $\Gamma'(0)/\Gamma(0)$ could arise from derivatives of gamma functions, but it is discarded in the limiting procedure as it is of the form
\begin{equation} \label{gammaterms}
    \frac{\Gamma(-iz / \alpha \beta_R) \Gamma'(iz/ \alpha \beta_R)}{\Gamma^2 (iz/ \alpha \beta_R)} + \frac{\Gamma(iz / \alpha \beta_R) \Gamma'(-iz/ \alpha \beta_R)}{\Gamma^2 (iz/ \alpha \beta_R)}.
\end{equation}
In fact, using that
\begin{equation}
    \Gamma(\epsilon) = - \Gamma(-\epsilon) \frac{\Gamma(1+\epsilon)}{\Gamma(1-\epsilon)}
\end{equation}
one concludes that \eqref{gammaterms} is zero. 

Furthermore
\begin{equation}
    2\pi i \text{Res}(f(z), \overline{0})= -\frac{i}{2} \sum_{n>0} \frac{(-1)^n}{(n!)^2} \frac{1}{e^{i \alpha \beta_R n}-1} \frac{\Gamma(n-s) \Gamma(1+s+n)}{\Gamma(1+s) \Gamma(-s)} e^{-2 |\lambda| \alpha n},
\end{equation}
which tends to zero as $\lambda \to -\infty$.

So in the limit when both points of the propagator are located on the horizon, we obtain
\begin{equation}
    \int_{-\infty}^{\infty} \frac{d \w }{4 \pi \w} \frac{\overrightarrow{A}_l (\w) e^{-2 i \lambda \w }}{e^{\beta_R \w}-1} \approx  \frac{|\lambda|}{2 \beta_R}, \qquad \lambda \to -\infty.
\end{equation}
The other complex conjugated term in \eqref{prop} gives the same contribution. Then, using that
\begin{align}
    P_l (\vec{x} \cdot \vec{y} ) = \frac{4 \pi}{2l+1} \sum_{m=-l}^l Y_{lm} (\vec{x}) Y_{lm}^*(\vec{y}), \qquad \sum_{l=0}^{\infty} \frac{2l+1}{2} P_l (x) P_l(y) = \delta(x-y),
\end{align}
the Wightman function can be rewritten as
\begin{equation} \label{schw_res}
    W(x,x') \approx  \frac{|\lambda|}{ \beta_R} \frac{\delta(\Omega,\Omega')}{(2M)^2} , \qquad \lambda \to - \infty.
\end{equation}
Actually, one can obtain the same answer without approximating the potential and employing exact solution in fitted potential. Namely, one can use the approximate near-horizon solution \eqref{near_horizon_mode},
but ultimately the result \eqref{schw_res} remains the same. However, the near-horizon approximation suffers from the fact that the potential for the radial function in this case is infinitely growing, so that one can only introduce the out-going modes that are fully reflected back from the potential, i.e. in this situation one has $|\overrightarrow{A}_l (\omega)|=1$. The approximation with the P{\"o}schl-Teller potential shows that deformation of the potential from the infinitely growing with no in-going modes to the bump-like with the presence of the transmitted in-going modes does not change the answer for the anomalous singularity.
This shows that this anomalous singularity is not a peculiarity for a given solution of Einstein's equations, but actually its presence does not depend on the form of the potential.

Also, one can see that for the discrete temperature $\beta = \frac{8 \pi M}{n}$, the result \eqref{schw_res} can be obtained from \eqref{wightman_discrete_sum} by taking the asymptotic form of the $K_0(x)$ as $x \to 0$. Dependence on $l$ splits off from the rest of the argument of this function because of the logarithm function in this asymptotic, so the leading terms in the horizon limit are given by the terms that depend on the geodesic distance, and the sum over $l$ gives the angular delta-function in \eqref{schw_res}.

\subsubsection{Massive case}

As the anomalous singularity on the horizon comes from the low-frequency modes (see previous subsection or \cite{Akhmedov:2020ryq}), we neglect the contribution from high-frequency modes in \eqref{mode_massive_sch}. Then,
the Wightman function in the horizon limit is as follows:
\begin{equation}
    W(x,x') \approx \sum_{lm} \int^{+\mu}_{-\mu} \frac{d \omega}{\pi \w}  \frac{e^{i \w (t-t')} R^*_l(\w|r) R_l (\w|r') }{e^{\beta_0 \w}-1} Y_{lm}(\Omega) Y^*_{lm}(\Omega')
\end{equation}
Assuming that $\mu \beta_0$ is sufficiently large for the integrand to be zero at the endpoints of the integration domain, we extend the domain of integration to the whole real axis, and the integral evaluated in complete analogy with the two-dimensional case \cite{Akhmedov:2020ryq}. Namely, in the limit \eqref{limit}, the Wightman function is written as
\begin{multline}
    W(x,x') \approx \sum_{lm} \frac{Y_{lm}(\Omega) Y_{lm}^*(\Omega')}{(2M)^2} \int^{\infty}_{-\infty} \frac{d\w}{4\pi \w} \frac{1}{e^{\beta_0 \w}-1} \bigg(e^{-i\w c} +  e^{2 i \w \lambda+2i \delta_{\w l}}
    + e^{i\w c} + e^{-2i \w \lambda-2i \delta_{\w l}} \bigg),
\end{multline}
where $\delta_{\w l}$ is given by \eqref{delta_sch}.
Evaluating the contribution of the double pole at the $\w=0$ as in the massless case, 
we obtain that
\begin{equation} \label{sing_schw_massive}
    W(x,x') \approx \frac{| \lambda |}{ \beta_0} \frac{\delta(\Omega,\Omega')}{(2M)^2}, \qquad \lambda \to -\infty,
\end{equation}
where $\delta(\Omega,\Omega')$ is the delta function on the two-sphere.

\section{Reissner-Nordström black hole} \label{RNsection}
In this section first we construct modes and Wightman functions outside of the outer horizon. We investigate behavior of the Wightman function on the outer horizon and evaluate the anomalous singularity when both points are located there. The resulting singularity is analogous to the one in the Schwarzschild black hole.

\subsection{Modes and Wightman function}
\begin{figure}[!h]
\centering
\includegraphics[width=0.6\textwidth]{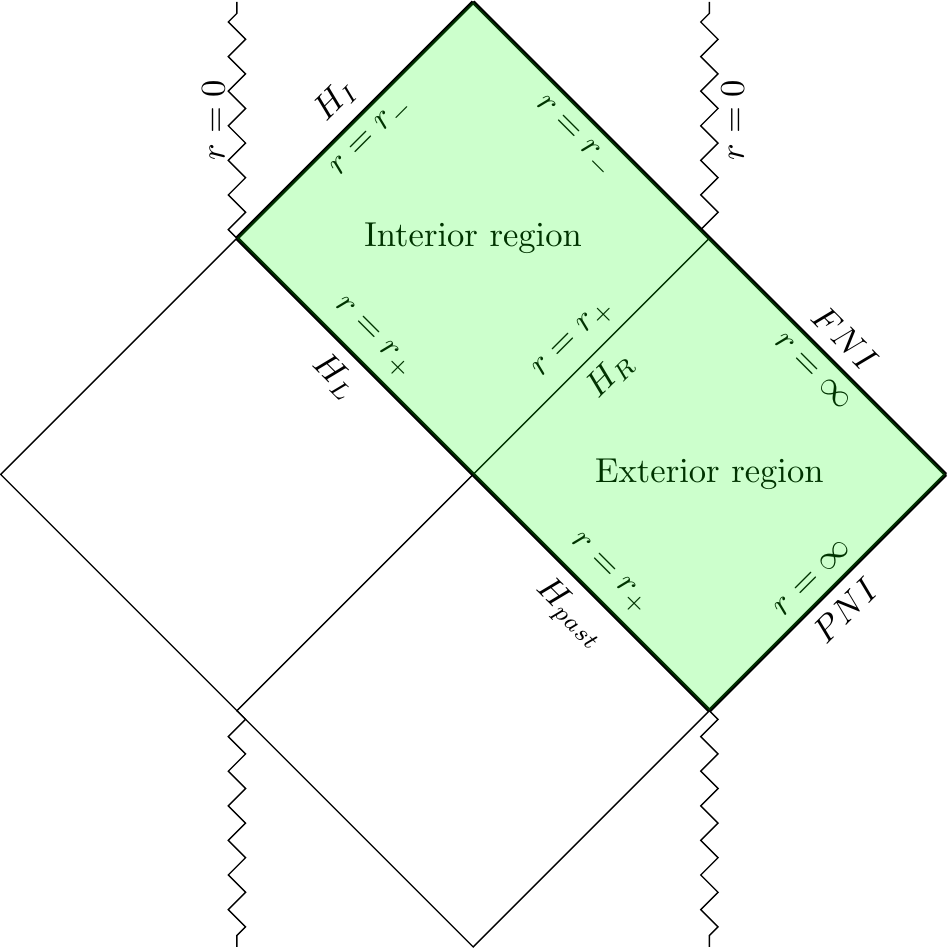}
\caption{Penrose diagram of the Reissner-Nordström black hole.} \label{picRN}
\end{figure}
Geometry of the Reissner-Nordström black hole is given by the metric
\begin{equation}
    ds^2 = \bigg(1-\frac{2M}{r}+\frac{Q^2}{r^2}\bigg) dt^2 - \frac{dr^2}{\big(1-\frac{2M}{r}+\frac{Q^2}{r^2}\big)}-r^2(d \theta^2 + \sin^2 \theta d \varphi^2).
\end{equation}
This geometry has two horizons at
\begin{equation}
    r_{\pm} = M \pm \sqrt{M^2-Q^2},
\end{equation}
where the surface $r=r_+$ is the event horizon and the surface $r=r_-$ is a Cauchy horizon (we consider non-extremal case $Q<M$). Surface gravity corresponding to these horizons is equal to: 
\begin{align}
    \kappa_{\pm} = \frac{r_+-r_-}{2r^2_{\pm}},
\end{align}
and the tortoise coordinate may be defined as
\begin{align}
    r_* = r+\frac{1}{2\kappa_+} \log \bigg( \frac{|r-r_+|}{r_+-r_-} \bigg) -\frac{1}{2\kappa_-} \log \bigg( \frac{|r-r_-|}{r_+-r_-} \bigg).
\end{align}
Also, it will be useful to define the Eddington-Finkelstein coordinates, which are written in the exterior region as
\begin{align}
    u = t-r_*, \qquad v = t+r_*,
\end{align}
In the following I consider the situation in which both points of the Wightman function are located on the outer horizon $r=r_+$, which is denoted as $H_R$ in Fig. \ref{picRN}.

\subsubsection{Massless case}

In the exterior region we can proceed analogously as in the Schwarzschild spacetime. Namely, we split variables as in \eqref{split}
and represent these functions in the form \eqref{f_split}, where the radial functions $R(r)$ satisfy the equation as follows
\begin{align} \label{exterior_radial}
    &\qquad \qquad \qquad  \frac{d^2R}{dr_*^2}+\big[ \w^2 - V_l (r) \big] R = 0, \nonumber \\
    &V_l(r) = \bigg(1-\frac{2M}{r}+\frac{Q^2}{r^2}\bigg) \bigg( \frac{l(l+1)}{r^2} + \frac{2M}{r^3} - \frac{2Q^2}{r^4} \bigg).
\end{align}
Again we distinguish two types of solutions \eqref{set1} of this equation with boundary conditions \eqref{boundary_cond}, corresponding to out-going and in-going modes. Then, as in the Schwarzschild spacetime, the Wightman function in the generic thermal state is written as
\begin{multline} \label{outer_hh}
    W(x,x') = \sum_{lm} \int^{+\infty}_{-\infty} \frac{d \omega}{4\pi \w} \frac{1}{rr'}\bigg[ \frac{e^{i \w (t-t')} \overrightarrow{R}^*_l (\w|r) \overrightarrow{R}_l (\w|r') }{e^{\beta_R  \w}-1}
  +\\+\frac{e^{i \w (t-t')} \overleftarrow{R}^*_l (\w|r) \overleftarrow{R}_l (\w|r') }{e^{\beta_L \w}-1} \bigg] Y_{lm} (\Omega) Y^*_{lm} (\Omega').
\end{multline}

\subsubsection{Massive case}
In the massive case the equation for the radial function is
\begin{align} \label{radial_rn}
    &\frac{d^2R}{dr_*^2}+\big[ \w^2 - V_l (r) \big] R = 0,
\end{align}
with the potential
\begin{equation} \label{potential}
    V_l (r) = \bigg( 1 - \frac{2M}{r} + \frac{Q^2}{r^2} \bigg) \bigg(  \frac{l(l+1)}{r^2} + \frac{2M}{r^3} - \frac{2Q^2}{r^4} + \mu^2 \bigg).
\end{equation}
Again, the modes with $\w^2 \leq \mu^2$ are
\begin{align}
    &\varphi_{\w lm} (x) = \frac{1}{\sqrt{ \pi |\w|}} e^{-i |\w| t} R_{ l} (\w | r) Y_{lm} (\Omega), \nonumber \\
    &R_{ l} (\w | r) \approx \frac{1}{r} \cos (\w r_* + \delta_{\w l}),
\end{align}
where the expression for the phase acquires the form
\begin{align}
    &\delta_{\w l} \approx \frac{\pi}{2} +\frac{\w}{2 \kappa_+} \log \big( \Tilde{l}^2+\mu^2 \big) -\w r_+ - \text{arg} \, \Gamma(1+\frac{i\w}{\kappa_+}), \nonumber \\
    &\qquad \qquad \qquad \qquad  \Tilde{l}^2 = l^2+l+1-\frac{r_-}{r_+}.
\end{align}
The modes with $\w^2 > \mu^2$ are defined as in \eqref{f_modes}. Then, the Wightman function is as follows:
\begin{multline}
    W(x,x') = \sum_{lm} \int^{+\mu}_{-\mu} \frac{d \omega}{\pi \w}  \frac{e^{i \w (t-t')} R^*_l(\w|r) R_l (\w|r') }{e^{\beta_0 \w}-1} Y_{lm}(\Omega) Y^*_{lm}(\Omega') + \\
    +\sum_{lm} \int_{|\w|>\mu} \frac{d \omega}{4\pi \w} \frac{1}{rr'} \bigg[ \frac{ e^{i \w (t-t')} \overrightarrow{F}^*_{ l}(\w | r) \overrightarrow{F}_{l}(\w|r')}{e^{\beta_R \w}-1} 
 +
  \frac{ e^{i \w (t-t')} \overleftarrow{F}^*_{l}(\w|r) \overleftarrow{F}_{l}(\w|r')}{e^{\beta_L \w}-1} \bigg] Y_{lm}(\Omega) Y^*_{lm}(\Omega') .
\end{multline}

\subsection{Anomalous singularity}

\subsubsection{Massless case}
As in the previous section, we take the near outer horizon limit $r \to r_+$ \eqref{limit} of the expression \eqref{outer_hh}, evaluate the reflection coefficients $\overrightarrow{A}_l (\w)$ by solving the radial equation near the outer horizon to obtain
\begin{align}
    &\overrightarrow{R}_l (\w |r) \approx 2 \Tilde{l}^{-\frac{i\w}{\kappa_+}}\frac{e^{i \w r_+}}{\Gamma(-i \w / \kappa_+)} K_{\frac{i\w}{\kappa_+}} \big( 2 \Tilde{l} e^{\kappa_+ (r_*-r_+)} \big), \nonumber \\ 
    &\overrightarrow{A}_l (\w) \approx - \frac{\kappa_+}{\pi \w} e^{- \frac{2i\w}{\kappa_+} \log \Tilde{l} +2i \w r_+} \sinh \bigg( \frac{\pi \w}{\kappa_+} \bigg) \Gamma \bigg( 1+\frac{i\w}{\kappa_+} \bigg).
\end{align}
Substituting this reflection coefficient into the expression for the Wightman function in the outer horizon limit
\begin{multline}
    W(x,x') \approx \frac{1}{(2M)^2} \sum_{lm} Y^*_{lm} (\theta, \varphi) Y_{lm} (\theta',\varphi')  \int_{-\infty}^{\infty} \frac{d \w }{4 \pi \w} \bigg[ \frac{1}{e^{\beta_R \w }-1} \big( e^{-i\w c} + \overrightarrow{A}_l (\w) e^{-2 i \lambda \w } + \\ + \overrightarrow{A}_l^* (\w) e^{2 i \lambda \w } + | \overrightarrow{A}_l (\w) |^2 e^{i\w c} \big) + \frac{1}{e^{\beta_L \w}-1} | B_l (\w) |^2 e^{i \w c} \bigg],
\end{multline}
and evaluating the integrals as in the Schwarzschild spacetime, one obtains
\begin{equation}
    W(x,x') \approx \frac{|\lambda| }{\beta_R} \, \frac{ \delta(\Omega,\Omega')}{r_+^2}   \, , \qquad \lambda \to -\infty.
\end{equation}

\subsubsection{Massive case}
As is the Schwarzschild spacetime, the leading contribution in the horizon limit \eqref{limit} comes from the low frequencies. Then, for the Wightman function in the horizon limit one has:
\begin{multline}
    W(x,x') \approx \sum_{lm} \int_{-\mu}^{\mu} d \w \, \frac{\varphi_{\w lm}^* (x) \, \varphi_{\w lm} (x')}{e^{\beta_0 \w}-1} \approx \\
    \approx \sum_{lm} \frac{Y_{lm}(\Omega) Y_{lm}^*(\Omega')}{(2M)^2} \int^{\infty}_{-\infty} \frac{d\w}{4\pi \w} \frac{e^{i \w (t-t')}}{e^{\beta_0 \w}-1} \bigg( e^{i \w (r_*+r_*')+2i \delta_{\w l}} + e^{-i \w (r_*+r_*')-2i \delta_{\w l}} \bigg).
\end{multline}
Evaluating the integrals as in the Schwarzschild spacetime, one obtains the result
\begin{equation}
    W(x,x') \approx \frac{| \lambda |}{ \beta_0} \frac{\delta(\Omega,\Omega')}{r_+^2}, \qquad \lambda \to -\infty.
\end{equation}
Thus, again we encounter the anomalous singularity at the outer horizon.

\section{Conclusions and discussions}

It is shown that the anomalous singularity also occurs in the four--dimensional Schwarzschild and Reissner-Nordström black hole backgrounds. 
Furthermore, as in Rindler and in de Sitter $d$--dimensional examples \cite{Akhmedov:2020ryq}, it is shown that anomalous singularity does not occur when the coordinates additional to $(t,r)$ do not coincide, and the divergence is amplified if they do. As we can represent delta function of zero argument as a linear divergence, we obtain that the power of the divergence is quadratic, as is expected from the analyisis in the Rindler and de Sitter spacetimes where in $d$ dimensions the divergence was of power $(d-2)$. These anomalous singularities lead to the explosive behavior of the expectation value of the stress-energy tensor, which means that with non-canonical temperature the backreaction of the QFT on the background geometry is strong. Does that mean that one cannot place a black hole in a bath with non-canonical temperature? This is the question for the further study.

One more possible outcome of this anomalous singularity is that it can affect the beta function as the coefficient before the light-like divergence is different than the canonical ones. 
It is interesting to see whether this anomalous singularity affects loop contributions to the two-point function in an interacting theory. Also, it is tempting to see how the results of this paper affect computation of the entropy of the fields outside the black hole. However, these problems are out of the scope of this paper and will be considered elsewhere.


\section{Acknowledgements}

I would like to thank E.T. Akhmedov, K.V. Bazarov, and D.V. Diakonov for valuable discussions.

This work was supported by the grant from the Foundation for the Advancement of Theoretical Physics and Mathematics “BASIS” and by Russian Ministry of education and science.

\end{document}